\documentclass{iaus}
\usepackage{graphicx}

\title[The Magnetic Field of HR 7355] 
{Discovery of a strong magnetic field in the \\ rapidly rotating B2Vn star HR 7355}

\author[M. E. Oksala et al.]   
{M.E. Oksala$^{1,2}$, G.A. Wade$^2$, W.L.F. Marcolino$^{3,4}$, J. Grunhut$^{2,5}$, \\
D. Bohlender$^{6}$, N. Manset$^{7}$, R.H.D. Townsend$^{8}$, and the MiMeS Collaboration}

\affiliation{$^{1}$Department of Physics and Astronomy,  University of Delaware, Newark, DE, USA\\[\affilskip]
$^2$Department of Physics, Royal Military College of Canada, Kingston, Ontario, Canada\\[\affilskip]
$^{3}$LAM-UMR 6110, CNRS \& Univ. de Provence, Marseille , France\\[\affilskip]
$^{4}$Observat\`{o}rio Nacional, Rio de Janeiro, Brazil  \\[\affilskip]
$^{5}$Department of Physics, Queen's University, Kingston, Ontario, Canada\\[\affilskip]
$^{6}$National Research Council of Canada, Herzberg Institute of Astrophysics,Victoria, Canada\\[\affilskip]
$^{7}$Canada-France-Hawaii Telescope Corporation,  Kamuela, HI, USA \\[\affilskip]
$^{8}$Department of Astronomy, University of Wisconsin-Madison, Madison, WI, USA}

\pubyear{2010}
\volume{272}  
\pagerange{1--2}
\jname{Active OB Stars: structure, evolution, mass-loss and critical limits}
\editors{C. Neiner, G. Wade, G. Meynet \& G. Peters, eds.}
\begin{document}

\maketitle

\begin{abstract}
We report on the detection of a strong, organized magnetic field in the helium-variable early B-type star HR 7355 using spectropolarimetric data obtained with ESPaDOnS on CFHT by the MiMeS large program. We also present results from new V-band differential photometry obtained with the CTIO 0.9m telescope. We investigate the longitudinal field, using a technique called Least-Squares Deconvolution (LSD), and the rotational period of HR~7355.  These new observations strongly support the proposal that HR 7355 harbors a structured magnetosphere similar to that in the prototypical helium-strong star, $\sigma$ Ori E.
\end{abstract}

\firstsection 

\section{Introduction}

HR~7355 (HD~182180) is a bright B2Vn helium-strong star originally classified as a Be star due to H$\alpha$ emission 
present in its spectrum (\cite[Abt \& Cardona 1984]{Abt_Cardona1984}).  Previous studies of this star show a $v \sin i \sim$ 300 km s$^{-1}$ 
(\cite[Abt et al. 2002]{Abt_etal2002}) with a P$_{\rm{rot}} \sim$ 0.52 d (\cite[Koen \& Eyer 2002]{Koen_Eyer2002}), 
as well as variation in helium, H$\alpha$, and brightness, suggesting the presence of a magnetosphere (\cite[Rivinius et al. 2008]{Rivinius_etal2008}).  
HR~7355 is the most rapidly rotating helium-strong star, rotating near its critical velocity, providing an excellent testbed for magnetospheres under the effects of rapid rotation.
 \vspace*{-0.5 cm}

\section{Method}

Least-Squares Deconvolution (LSD) describes the stellar spectrum as the
convolution of a mean Stokes I or V profile, representative of the average
shape of the line profile, and a line mask, describing the position, strength
and magnetic sensitivity of all lines in the spectrum.  From the LSD mean Stokes I and V profiles, we calculate the
longitudinal magnetic field, B$_{\ell}$:
\begin{equation}
B_{\ell} = -2.14 \times 10^{11} \frac{ \int v V(v) dv}{\lambda g c \int [1-I(v)]dv}
\end{equation}
 (\cite[Wade et al. 2000]{Wade_etal2000}), where $\lambda$ is the average wavelength and g is the average Land\'e factor in the mask. I$_{c}$ is
the continuum value of the intensity profile. The integral is evaluated over the full
velocity range of the mean profile.  

 \vspace*{-0.25 cm}

\section{Results}

We detect a strong magnetic field on HR~7355, the most rapidly rotating helium-strong star discovered thus far.  A simultaneous independent confirmation of the field detection has been obtained with FORS at the VLT by Rivinius et al. 2010.  The longitudinal magnetic field varies sinusoidally with the rotation period, with extrema -2 to 2.5 kG. 
Assuming a dipole magnetic field, the polar value of the magnetic field is $\sim$ 13-17 kG.  The photometric (brightness) light curve constructed from HIPPARCOS archival data and new CTIO measurements shows two minima separated by 0.5 in rotational phase and occurring 0.25 cycles before/after the magnetic extrema. Using the Scargle periodogram, eclipse-like photometric variations give a highly precise P$_{rot}$ = 0.5214404(6) days. We confirm spectral variability of helium and metal lines, as well as variability of H$\alpha$ emission. H$\alpha$ emission indicates circumstellar material extending out to 5 R$_{\star}$ from the star, rotating rigidly with the stellar surface.
We conclude that HR 7355 is a magnetic oblique rotator with a magnetosphere, mirroring the
physical picture for $\sigma$ Ori E (\cite[Townsend et al. 2005]{Townsedn_etal2008}).  

\begin{figure}[ht]
\begin{center}
 \includegraphics[width=3.4in]{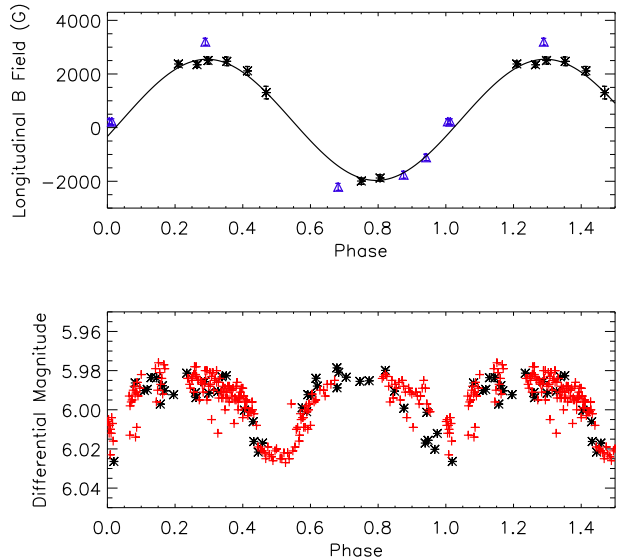} 
 \caption{ Top: Longitudinal magnetic field measurements for HR 7355 and the best-fit first order
sine curve. \cite[Oksala et al. (2010)]{Oksala_etal2010} (asterisks) and \cite[Rivinius et al. (2010)]{Rivinius_etal2010}  (diamonds) with 1 $\sigma$ error bars. 
 Bottom: The V-band photometric light curve for HR 7355 including both  HIPPARCOS photometry (asterisks) and new CTIO data (plus signs).}
   \label{fig1}
\end{center}
\end{figure}

\end{document}